\documentclass[prl,aps]{revtex4}

\usepackage{amssymb}
\usepackage{amsmath}
\usepackage{epsf}
\usepackage{latexsym}
\usepackage{psfrag}
\usepackage{epsfig}
\usepackage{bm}

\newcommand{\ord}{{\cal O}}

\newcommand{\nn}{\nonumber\\} 

\newcommand{\f}[1]{\vec #1}

\newcommand{\vau}{\vec v}
\newcommand{\na}{\vec\nabla}
\newcommand{\bea}{\begin{eqnarray}}
\newcommand{\ea}{\end{eqnarray}}

\begin{document}

\title{Quantum back-reaction problems\footnote{Proceedings to the
    workshop {\em From Quantum to Emergent Gravity: Theory and
    Phenomenology}, June 11-15 2007, Trieste, Italy, published in
    Proceedings of Science}}


\author{Ralf Sch\"utzhold\footnote{E-mail: schuetz@theory.phy.tu-dresden.de}}

\affiliation{Institut f\"ur Theoretische Physik,
Technische Universit\"at Dresden, D-01062 Dresden, Germany}

\begin{abstract}
The macroscopic behavior of many physical systems can be
  approximately described by classical quantities. However, quantum
  theory demands the existence of omnipresent quantum fluctuations on
  top of this classical background -- which, albeit small, should have
  some impact onto its dynamics. The correct treatment of this quantum 
  back-reaction is one of the main problems in quantum gravity and 
  related to fundamental questions such as the initial (big bang)
  singularity or the cosmological constant. By means of the
  qualitative analogy between gravity and fluid dynamics, we try to
  shed some light onto these problems and show some of the
  difficulties associated with the calculation of the quantum
  back-reaction starting from the classical (macroscopic) equation of
  motion.
\end{abstract}

\maketitle



\section{Motivation}

In physics (and other sciences), it is often useful to make 
{\em gedanken} experiments in order to explore the limiting cases of a
given theory and to approach it from different sides. 
Therefore, let us do the same here and imagine the following
situation: 
While sitting on the coffee table in your physics department, 
a colleague approaches you to ask you a favor:
``Hi, I have a question for you -- there is this theory with the
following properties:''   
\begin{itemize}
\item 
The classical equations of motion are known and my experimental
collaborators have them reasonably well tested at large length scales
and low energies.   
\item
However, we all agree that this classical description should break
down at small length scales, where somehow quantum effects should
become important. 
\item
But if I try to quantize this theory in the usual perturbative manner,
I get non-renormalizable UV-divergences and apparently I have to
introduce new (unknown) parameters.   
\item
Unfortunately, I do not know and understand the the full quantum
theory (yet). 
\end{itemize}
``Nevertheless, my experimental collaborators are now preparing precision
measurements and they want me to estimate the back-reaction of the
quantum fluctuations onto the classical background profile 
(as described by the classical equations of motion).
Do you have any idea how to do this?''

\section{Fluid Dynamics}

From reading the above points, one might have expected that this
colleague works in quantum gravity -- but in fact, precisely the same
problem does also apply to fluids dynamics:
\begin{itemize}
\item
The classical description is provided by the Euler equation 
\bea
\label{Euler}
\frac{d}{dt}\vau=
\left(\frac{\partial}{\partial t}
+\vau\cdot\na\right)\vau=
-\frac{\na p}{\varrho}+\frac{\f{f}_{\rm ext}}{\varrho}
\,,
\ea
together with the equation of continuity 
$\dot\rho+\na\cdot(\rho\vau)=0$ plus the equation of state $p=p(\rho)$
and the external force density $\f{f}_{\rm ext}$. 
\item
However, this equation clearly breaks down at small length scales,
where the continuum fluid picture fails. 
\item
The perturbative quantization of the longitudinal degrees of freedom
yields phonons as quasi-particles and the inclusion of phonon loops
produces non-renormalizable UV-divergences. 
Moreover, the quantization of the transversal degrees of freedom,
i.e., the vorticity $\na\times\vau$, cannot be inferred from the 
Euler equation at all.
E.g., in superfluids, one can only have an integer number of vortices
and the circulation quantum of a single vortex depends on the mass $m$
of the constituent particles -- which is not contained in Euler
equation.  
\item
Even in those situations where the the full quantum theory 
(i.e., the many-particle Hamiltonian) is in principle known, 
it is practically impossible to solve -- except in a few simple
cases.
(We shall study one of them below, see also \cite{back}.)
\end{itemize}
%

\section{First Try}

Clearly, the phonon modes derived from Eq.~(\ref{Euler}) must also
possess quantum fluctuations and these quantum fluctuations should
somehow react back onto the fluid.  
Let us try to estimate this impact.
In our first attempt, we assume zero knowledge of the microscopic
structure of the fluid and just use the classical equation of motion. 
In order to avoid the aforementioned problems with vorticity, we
assume an irrotational flow as described by the Bernoulli equation 
\bea
\label{Bernoulli}
\frac{\partial}{\partial t}\phi
+
\frac12(\na\phi)^2
+V_{\rm ext}
=-h(\varrho)=-g\varrho
\,,
\ea
with $\phi$ being the velocity potential $\vau=\na\phi$.
The specific enthalpy $dh=dp/\varrho$ determines the equation of
state, where we have chosen the relation $h(\varrho)=g\varrho$
describing Bose-Einstein condensates for later comparison. 

Now the usual procedure consists in replacing the c-number fields
$\rho$ and $\phi$ in the equation above by operators (more precisely,
operator-valued distributions) and to interpret their expectation
values as the classical contribution (mean-field expansion) 
\bea
\label{mean-field}
\phi &\to& \hat\phi
=
\langle\hat\phi\rangle+\delta\hat\phi
=\phi_{\rm cl}+\delta\hat\phi
\nn
\varrho &\to& \hat\varrho
=
\langle\hat\varrho\rangle+\delta\hat\varrho
=\varrho_{\rm cl}+\delta\hat\varrho
\,.
\ea
Inserting the above split into the Bernoulli equation
(\ref{Bernoulli}) and taking the expectation value yields 
\bea
\dot\phi_{\rm cl}
+
\frac12(\na\phi_{\rm cl})^2
+V_{\rm ext}
+g\varrho_{\rm cl}
=
-\frac12\langle(\na\delta\hat\phi)^2\rangle
\,,
\ea
where the last term can be attributed to the quantum fluctuations of
the phonon modes and should include their zero-point pressure. 
The quantum fluctuations $\delta\hat\phi$ of sound derived from
Eq.~(\ref{Bernoulli}) behave exactly like
a scalar (Klein-Fock-Gordon) quantum field -- provided that one
replaces the speed of light with the sound velocity $c_s$ in all 
expressions. 
Hence the expectation value in the above equation is UV-divergent 
and behaves as $\int k\,d^3k$. 
Remembering that the fluid picture is only valid for large length
scales, one would cut off the $k$-integration at some maximum
wavenumber $k_{\rm cut}$.
Below this cut-off $k_{\rm cut}$, our description should work fairly
well and hence one would arrive at the following estimate 
(see also \cite{balbinot}) 
\bea
\label{first-try}
p_{\rm zero-point}
=
-\frac12\langle(\na\delta\hat\phi)^2\rangle
\propto
k_{\rm cut}^4
\,.
\ea
An analogous line of reasoning has been adopted for estimates of the
cosmological constant (i.e., the gravitational impact of the quantum
vacuum) $\Lambda\propto M_{\rm Planck}^4$, which turns out to be many
orders of magnitude too large.  
However, in the above approach the main contribution to the zero-point 
pressure stems from phonon modes with wavenumbers close to 
$k_{\rm cut}$, where the theory breaks down.
Therefore, one might worry whether the above estimate is reliable. 

\section{Second Try}

In order to tackle this question, let us make a second attempt by
incorporating some knowledge about what is going on at high
wavenumbers $k_{\rm cut}$, i.e., how the fluid picture changes at
small length scales. 
Of course, this behavior will depend on the kind of fluid under
consideration -- i.e., we have to specify it. 
In the following, we shall consider dilute Bose-Einstein condensates,
since we understand the microscopic structure of these quantum fluids
quite well. 
In Bose-Einstein condensates, the Bernoulli equation breaks down --
i.e., acquires additional terms -- at the healing length
$\xi=1/\sqrt{mg\varrho}$ ($\hbar=1$ throughout)
\bea
\label{Bernoulli-BEC}
\dot\phi
+
\frac12(\na\phi)^2
+V_{\rm ext}
+g\varrho
=
\frac{1}{2m}\,\frac{\na^2\sqrt{\varrho}}{\sqrt{\varrho}}
\,.
\ea
For large length scales $k\ll1/\xi$, the last term\footnote{For
  historical reasons, this term is called ``quantum pressure'', but
  that nomenclature is a bit misleading since this term already occurs
  on the classical level -- e.g., in Eq.~(\ref{Bernoulli-BEC}), which
  is purely classical.} 
can be neglected and we recover the usual sound waves with a linear
dispersion relation $\omega=c_sk$.
For higher wavenumbers, however, the dispersion relation changes and
approaches the energy-momentum relation $\omega=k^2/(2m)$ for free
particles (with mass $m$) in the limit $k\xi\gg1$
\bea
\label{dispersion}
\omega^2=c_s^2k^2+\frac{k^4}{4m^2}
\,.
\ea
If we now insert the same mean-field expansion (\ref{mean-field}) as
in the previous Section into the modified Bernoulli equation
(\ref{Bernoulli-BEC}) and neglect terms of third or higher order in
$\delta\hat\varrho$, we get 
\bea
p_{\rm zero-point}
=
-\frac12\langle(\na\delta\hat\phi)^2\rangle
+\frac{1}{8\varrho^2}\langle(\na\delta\hat\varrho)^2\rangle
+\ord(\delta\hat\varrho^3)
\,.
\ea
Without respecting the altered dispersion relation (\ref{dispersion}),
the additional term $\langle(\na\delta\hat\varrho)^2\rangle$ would be
even stronger UV-divergent $\int k^3\,d^3k$ than the first one 
$\langle(\na\delta\hat\phi)^2\rangle$. 
However, including the full dispersion relation (\ref{dispersion}),
both expectation values behave as $\int k^2\,d^3k$ for large $k$ and
the leading UV-divergences cancel exactly.
Nevertheless, the sub-leading UV-singularities remain and one obtains 
\bea
\label{second}
p_{\rm zero-point}
\propto
gk_{\rm cut}^3
\,.
\ea
Therefore, the $k_{\rm cut}^4$ behavior in estimate (\ref{first-try})
we just ``successfully'' calculated in the previous Section is
completely canceled by the additional term in
Eq.~(\ref{Bernoulli-BEC}), which becomes relevant for wavenumbers
$k=\ord(1/\xi)$ where the fluid picture breaks down. 
Furthermore, even after fully incorporating the cut-off 
$k_{\rm cut}^\xi\sim1/\xi$ stemming from the healing length, the
result is still UV-divergent (though the singularity is weaker now). 
Naturally, one would expect that this remaining divergence 
$k_{\rm cut}^3$ reflects the omitted details of the microscopic
structure of two-particle interaction potential which generates the
internal pressure in the condensate, cf.~\cite{mean}. 

\section{Third Try -- Full Theory}

Fortunately, for dilute Bose-Einstein condensates, we are able to
address this question analytically.
The full many-particle Hamiltonian describing the condensate reads 
\bea
\label{full}
\hat{H}=
\int d^3r
\left[
\frac{(\na\hat\Psi^\dagger)\cdot(\na\hat\Psi)}{2m}
+V_{\rm ext}\hat\Psi^\dagger\hat\Psi
\right]
+
\int d^3r\,d^3r'\,
\hat\Psi^\dagger(\f{r})\hat\Psi^\dagger(\f{r}\,')
V_{\rm int}(\f{r}-\f{r}\,')
\hat\Psi(\f{r})\hat\Psi(\f{r}\,')
\,,
\ea
where $V_{\rm int}(\f{r}-\f{r}\,')$ denotes the two-particle
interaction potential.  
In the $s$-wave approximation, the microscopic structure of this
potential is omitted and the Hamiltonian simplifies to 
\bea
\label{s-wave}
\hat{H}=
\int d^3r
\left[
\frac{(\na\hat\Psi^\dagger)\cdot(\na\hat\Psi)}{2m}
+V_{\rm ext}\hat\Psi^\dagger\hat\Psi
+\frac{g}{2}(\hat\Psi^\dagger)^2\hat\Psi^2
\right]
\,.
\ea
In order to make contact to the results of the previous Sections, we
introduce the density $\hat\varrho=\hat\Psi^\dagger\hat\Psi$ and phase
$\hat\phi$ operators via the quantum Madelung ansatz 
$\hat\Psi=e^{i\hat\phi}\,\sqrt{\hat\varrho}$.
Note, however, that this split is highly singular due to the
commutator $[\hat\Psi(\f{r}),\hat\Psi^\dagger(\f{r}\,')]=
\delta^3(\f{r}-\f{r}\,')$. 
Therefore, if we try to re-express the above Hamiltonian in terms of
$\hat\varrho$ and $\hat\phi$ 
\bea
(\hat\Psi^\dagger)^2\hat\Psi^2
=
(\hat\Psi^\dagger\hat\Psi)^2
-\hat\Psi^\dagger[\hat\Psi,\hat\Psi^\dagger]\hat\Psi
=
\hat\varrho^2-\hat\varrho\delta^3(0)
\,,
\ea
we encounter UV-divergent operator-ordering remnants 
$\delta^3(0)\propto k_{\rm cut}^3$. 
Thus the quantum analogue of the Bernoulli equation 
\bea
\frac{\partial\hat\phi}{\partial t}
=
i[\hat H,\hat\phi]
=
\frac{\dots}{2m}
-V_{\rm ext}
-g\hat\varrho
+\frac{g}{2}\delta^3(0)
\,,
\ea
contains at least one singular term 
$g\delta^3(0)\propto gk_{\rm cut}^3$. 
Comparison with Eq.~(\ref{second}) shows that the estimate of 
$p_{\rm zero-point}$ in the previous Section is completely spoilt by
such unphysical operator-ordering remnants. 
One might argue that a constant (though infinite) pressure shift does
not play any role -- but the coupling constant $g$ may well depend on
an external magnetic field and thereby vary spatially (which would
feign a pressure gradient). 

Note that the singularity $g\delta^3(0)\propto gk_{\rm cut}^3$
could be cured by a smooth two-particle interaction potential 
$V_{\rm int}(\f{r}-\f{r}\,')$.
However, insertion of the quantum Madelung split 
$\hat\Psi=e^{i\hat\phi}\,\sqrt{\hat\varrho}$ into the kinetic term 
$(\na\hat\Psi^\dagger)\cdot(\na\hat\Psi)$ 
again yields operator-ordering singularities which are independent of  
$V_{\rm int}(\f{r}-\f{r}\,')$ and thus the quantum
Bernoulli equation above is still ill-defined. 
Therefore, in order to calculate the back-reaction of the quantum
fluctuations correctly, one should not use the ill-defined phase
operator $\hat\phi$ but employ the fundamental field operators 
$\hat\Psi^\dagger$ and $\hat\Psi$ instead.
In analogy to (\ref{mean-field}), the mean-field split 
$\hat\Psi=\psi_{\rm c}+\hat\chi$ into the condensate order parameter
$\psi_{\rm c}$ plus small quantum fluctuations (phonon modes)
$\hat\chi$ yields the Gross-Pitaevskii equation 
\bea
i\dot\psi_{\rm c}
=
\left[
-\frac{\na^2}{2m}+V_{\rm ext}+g|\psi_{\rm c}|^2
+2g\left\langle\hat\chi^\dagger\hat\chi\right\rangle
\right]\psi_{\rm c}+
g\left\langle\hat\chi^2\right\rangle\psi_{\rm c}^*
\ea
including the quantum back-reaction terms 
$\left\langle\hat\chi^\dagger\hat\chi\right\rangle$
and 
$\left\langle\hat\chi^2\right\rangle$.
The first term $\left\langle\hat\chi^\dagger\hat\chi\right\rangle$ is
called the quantum depletion and is finite even in the $s$-wave
approximation 
$V_{\rm int}(\f{r}-\f{r}\,')\to g\delta^3(\f{r}-\f{r}\,')$.  
The second term $\left\langle\hat\chi^2\right\rangle$ 
(known as the anomalous contribution) would be UV-divergent in this 
approximation 
$V_{\rm int}(\f{r}-\f{r}\,')\to g\delta^3(\f{r}-\f{r}\,')$, but only
weakly $\left\langle\hat\chi^2\right\rangle\propto k_{\rm cut}$. 
Taking into account the finite range of the interaction 
$V_{\rm int}(\f{r}-\f{r}\,')\neq g\delta^3(\f{r}-\f{r}\,')$, however,
it is also finite. 
Therefore, the description in terms of the fundamental field operators 
$\hat\Psi^\dagger$ and $\hat\Psi$ is well-defined and finite --
whereas the phase operator $\hat\phi$ is ill-defined and generates
unphysical singularities. 

\section{Conclusions}

The presented {\em gedanken} experiment is based on a qualitative
analogy between (quantum) fluids and (quantum) gravity sketched in the
following table. 
Even though the problem of quantizing gravity is roughly as old as the
quest for a full quantum description of fluids, our progress in the
latter case is far more explicit (see the various question marks below). 
This difference can probably be mostly attributed to the vastly
different input from the experimental side.

\bigskip
\begin{tabular}{lccc}
Classical equation of motion: & 
Euler equation (\ref{Euler}) & $\leftrightarrow$ &
Einstein equations 
\\
Quantized linearizations: 
& Phonons $\hat\chi$ & $\leftrightarrow$ & 
Gravitons (?) 
\\
First UV cut-off scale: 
& Healing length $\xi$ & $\leftrightarrow$ & 
Planck scale $M_{\rm Planck}$?
\\
Full quantum theory: 
& Many-body Hamiltonian (\ref{full}) & $\leftrightarrow$ & 
Quantum gravity?
\\
Quantum back-reaction: 
& Zero-point pressure $p_{\rm zero-point}$ & $\leftrightarrow$ & 
Cosmological constant $\Lambda$?
\end{tabular}
\bigskip
\\
Now, what can we learn from our {\em gedanken} experiment, i.e., 
what are possible lessons for quantum gravity (based on our superior
understanding of quantum fluids)?
\begin{itemize}
\item
Usually, quasi-particles are obtained by quantizing the linearized
fluctuations of the classical equation of motion. 
For phonons, this procedure indeed yields the correct description for
low energies (on the linear level). 
However, the vorticity cannot be quantized in this way and requires
some knowledge of the microscopic structure of the fluid -- which is
not contained in the Euler equation. 
Whether this linearized quantization procedure works for gravitons is
not completely clear.
One might expect that they behave like Goldstone modes and thus are
similar to phonons --  but without knowing the full quantum theory,
this is not evident. 
\item
Going beyond linear order, one encounters (non-renormalizable) UV
divergences in both cases. 
As we have observed in the previous Sections, naively summing the
zero-point fluctuations of the phonon modes up to some cut-off 
(and thereby extrapolating the Euler equation to large $k$) does not 
give the correct result. 
This observation provides a different view on the naive estimate 
$\Lambda\propto M_{\rm Planck}^4$ of the cosmological constant,
cf.~\cite{volovik}. 
\item
A third point is the existence of two (or more) vastly different UV 
cut-off scales.
In Bose-Einstein condensates, the first deviation from the Bernoulli
equation occurs for length scales at the healing length 
$k_{\rm cut}^\xi\sim1/\xi$, where the dispersion relation changes. 
In gravity, a change of the dispersion relation corresponds to the
breakdown or modification of Lorentz invariance. 
However, even including the modified dispersion relation
(\ref{dispersion}), there are still UV divergences in the $s$-wave
scattering approximation (\ref{s-wave}), which can be removed by
taking into account the full two-particle interaction potential 
$V_{\rm int}(\f{r}-\f{r}\,')$.
Therefore, the range of $V_{\rm int}(\f{r}-\f{r}\,')$ introduces
another UV cut-off $k_{\rm cut}^g$ which is 
(in dilute Bose-Einstein condensates) much larger 
$k_{\rm cut}^g\gg k_{\rm cut}^\xi\sim1/\xi$ 
than the wavenumbers at which the effective Lorentz invariance breaks
down.
This observation poses the question of the significance of the UV
scales in quantum gravity -- does really everything happen at the
Planck scale and is there indeed nothing going on at higher scales?
\item
In dilute Bose-Einstein condensates with a pure contact interaction,
the first cut-off $k_{\rm cut}^\xi\sim1/\xi=\sqrt{mg\varrho}$ 
is directly related to the circulation quantum 
\bea
\Gamma=\oint d\vec r\cdot\vau=\frac{2\pi}{m}\,{\mathbb N}=
2\pi\xi c_s\,{\mathbb N}
\,,
\ea
which determines the ``quantization'' of vorticity. 
In other superfluids, however, the two scales can be independent.
In the presence of finite-range (e.g., dipolar) interactions, for
example, the first deviation of the dispersion relation is determined
by the maxon hill/roton dip, which depends on the interaction
structure -- while the circulation quantum is still determined by the
mass $m$.
(I.e., in this case, there are {\em three} UV scales.) 
\item
Finally, the presented calculations clearly demonstrate the importance
of identifying the correct fundamental variables.
Given a classical equation valid at large scales, one can generate
almost every outcome for the impact of the quantum fluctuations by
quantizing it in terms of different variables and using the operator 
ordering ambiguities.
In fact, the power of fluid dynamics partly relies on the fact that
many systems which are totally different on the microscopic level obey 
very similar classical equation of motions at large length scales and
low energies.
Therefore, it is crucial to identify universal (emergent) features and
to distinguish them from system specific properties. 
\end{itemize}

\section*{Acknowledgments}

The author acknowledges valuable discussions with Ted Jacobson, Renaud
Parentani, Bill Unruh, Grisha Volovik, and many others at the workshop
{\em From Quantum to Emergent Gravity: Theory and Phenomenology}
(SISSA, Trieste, Italy 2007) as well as support by the European
Science Foundation network programme 
``Quantum Geometry and Quantum Gravity''.  
This work was supported by the Emmy-Noether Programme of the German
Research Foundation (DFG, SCHU~1557/1-2). 



\begin{thebibliography}{99}

\bibitem{back} 
R.~Sch\"utzhold, M.~Uhlmann, Y.~Xu, and U.~R.~Fischer,
\emph{Quantum back-reaction in dilute Bose-Einstein condensates}, 
Phys.\ Rev.\ D {\bf 72}, 105005 (2005). 

\bibitem{balbinot}
R.~Balbinot, S.~Fagnocchi, A.~Fabbri and G.~P.~Procopio,
\emph{Backreaction in acoustic black holes},
Phys.\ Rev.\ Lett.\  {\bf 94}, 161302 (2005);
%
R.~Balbinot, S.~Fagnocchi and A.~Fabbri,
\emph{Quantum effects in acoustic black holes: The backreaction}, 
Phys.\ Rev.\  D {\bf 71}, 064019 (2005).

\bibitem{mean} 
R.~Sch\"utzhold, M.~Uhlmann, Y.~Xu and U.~R.~Fischer,
\emph{Mean-field expansion in Bose-Einstein condensates
with finite-range interactions}, 
Int.\ J.\ Mod.\ Phys.\ B {\bf 20}, 3555 (2006).

\bibitem{volovik}  
G.~E.~Volovik,
\emph{From Quantum Hydrodynamics to Quantum Gravity},
{\tt arXiv:gr-qc/0612134}, 
rapporteur article for 
session ``Analog Models of and for General Relativity'' in the 
Proceedings of the Eleventh Marcel Grossmann Meeting on General
Relativity, edited by H.~Kleinert, R.~T.~Jantzen and R.~Ruffini 
(World Scientific, Singapore, 2007). 

\end{thebibliography}
\end{document}